\documentclass[12pt, onecolumn]{IEEEtran}
\usepackage{amsfonts}
\usepackage{amssymb}
\usepackage{graphicx}
\usepackage{amsmath}
\usepackage{amsmath, amsfonts,epsfig, multirow}
\usepackage{epsfig,graphics,graphicx,latexsym,amsfonts,amssymb,amsmath,verbatim,slashbox}
\usepackage{subfigure}
\usepackage{cite}
\usepackage{color}
\usepackage{soul}
\usepackage{slashbox}
\usepackage{changebar}
\usepackage{mathrsfs}
\usepackage{epstopdf}

\graphicspath{{figures/}}

\begin{document}
\baselineskip 24pt
\parskip 9pt
\thispagestyle{empty}

\vspace*{2.0cm}
\begin{center}
{\LARGE{Soft-Defined Heterogeneous Vehicular Network: Architecture and Challenges}}
\end{center}

\vspace*{1cm}
\begin{center}

{\normalsize Kan Zheng {\it Senior Member,~IEEE}$^\ast$, Lu Hou$^\ast$, Hanlin Meng$^\ast$, Qiang Zheng$^\ast$, Ning Lu$^\ddag$, and Lei~Lei$^\dag$ {\it Member,~IEEE}\\

\vspace{0.3cm}

\small $^\ast$Wireless Signal Processing and Network Lab\\
Key laboratory of Universal Wireless Communication, Ministry of Education \\
Beijing University of Posts \& Telecommunications\\
Beijing, China\\

\small $^\ddag$ Department of Electrical and Computer Engineering\\
University of Waterloo\\
Waterloo, Ontario, Canada~~N2L 3G1\\

\small $^\dag$  State Key Laboratory of Rail Traffic Control \& Safety\\
Beijing Jiaotong University\\
Beijing, China, 100044\\

E-mail: \tt{kzheng@ieee.org} \\
}
\end{center}

\vspace{0.8cm} \hspace*{0.3in} \noindent {\normalsize {\bf
Correspondence:}~~Dr.~Kan Zheng \\
\hspace*{1.8in} Wireless Signal Processing and Network Lab \\
\hspace*{1.8in} Beijing University of Posts \& Telecommunications\\
\hspace*{1.8in} P.O.Box 93, No. 10, Xi Tu Cheng Road, Beijing, China   \\
\hspace*{1.8in} Phone: (+86) 10-61198066, ~Fax: (+86) 10-61198078\\
\hspace*{1.8in} Email: \hspace*{0.025in} zkan@bupt.edu.cn }

\newpage
\baselineskip 19pt
\parskip 4pt

\setcounter{page}{1}

\begin{center}

\vspace*{0mm}

{\LARGE \bf  Soft-Defined Heterogeneous Vehicular Network: Architecture and Challenges}

\vspace*{10mm}

{\normalsize Kan Zheng {\it Senior Member,~IEEE}$^\ast$, Lu Hou$^\ast$, Hanlin Meng$^\ast$, Qiang Zheng$^\ast$, Ning Lu$^\ddag$, and Lei~Lei$^\dag$ {\it Member,~IEEE}
\\

\vspace{0.3cm}

\small $^\ast$ Wireless Signal Processing and Network Lab \\
Key Lab of Universal Wireless Communications, Ministry of Education\\
Beijing University of Posts \& Telecommunications\\
Beijing, China, 100088\\

$^\ddag$ Department of Electrical and Computer Engineering\\
University of Waterloo\\
Waterloo, Ontario, Canada~~N2L 3G1\\

$^\dag$  State Key Laboratory of Rail Traffic Control \& Safety\\
Beijing Jiaotong University\\
Beijing, China, 100044\\

Contact email: kzheng@ieee.org \\

}
\end{center}

\vspace*{10mm}
\begin{center} {\bf Abstract}
\end{center}
Heterogeneous Vehicular NETworks (HetVNETs) can meet various quality-of-service (QoS) requirements for intelligent transport system (ITS) services by integrating different access networks coherently. However, the current network architecture for HetVNET cannot efficiently deal with the increasing demands of rapidly changing network landscape. Thanks to the centralization and flexibility of the cloud radio access network (Cloud-RAN), soft-defined networking (SDN) can conveniently be applied to support the dynamic nature of future HetVNET functions and various applications while reducing the operating costs. In this paper, we first propose the multi-layer Cloud RAN architecture for implementing the new network, where the multi-domain resources can be exploited as needed for vehicle users. Then, the high-level design of soft-defined HetVNET is presented in detail. Finally, we briefly discuss key challenges and solutions for this new network, corroborating its feasibility in the emerging fifth-generation (5G) era.

\par

\begin{flushleft}
\textbf{\textit{Index Terms}}-- Cloud-RAN, Soft-Defined Networking (SDN), Vehicular Communication.
\end{flushleft}

\newpage
\baselineskip 24pt
\parskip 9pt

\section{Introduction}
\label{SEC_INTRO}

Vehicular Networks have attracted lots of interests in the past decade from both industry and academia. Via enabling vehicles to communicate with each other, i.e., vehicle-to-vehicle(V2V), and with infrastructure, i.e., vehicle-to-infrastructure(V2I), vehicular networks can support enormous applications, which are mainly divided into two categories, i.e., safety applications and non-safety applications. Dedicated Short Range Communication (DSRC) based on IEEE 802.11p has gained increasing attention for its easy deployment, low cost and the capacity to naturally support V2V communications~\cite{DSRC}. Nonetheless, it suffers from scalability issues, e.g., unbounded delay, limited radio range and lacks of a pervasive roadside communication infrastructure~\cite{dtn1}. Thus, the forth generation (4G) Long Term Evolution (LTE) system is proposed to support the vehicular communication~\cite{LTEV}. For the vehicle user, it can benefit from the large coverage, high throughput and low latency of LTE networks. However, due to high vehicle mobility and dynamic network topology, it is quite difficult to provide satisfied ITS services only through LTE. Especially, when the number of vehicle users increases in the cell coverage, the strict latency requirement of safety application cannot be guaranteed by LTE itself. Heterogeneous Vehicular NETworks (HetVNET), by integrating different access networks including DSRC and LTE, is expected to be a viable solution to meet various communication requirements for ITS services~\cite{HET}. \par

HetVNETs are in the process of constituting a fundamental information platform, and will eventually evolve into all vehicles connected. However, there are problems that hinder the rapid development of HetVNET. ITS services proliferate significantly and different kinds of them can be well supported by different wireless access technology. Various heterogeneous wireless access technologies exist in HetVNET, which can hardly be well cooperated under the traditional network architecture. Then, a plenty of wireless network infrastructures and spectrum resources may be wasted and thereby lead to the low quality of experience (QoE) of vehicle users. This situation may become worse with the increase of network scale. It may take a long time for network operators to deploy new services since they need to configure or modify a large amount of underlying devices. All of these issues above call for rethink of the current wireless network architecture. More open and efficient architectures are expected to emerge. \par

Nowadays, cloud radio access network (Could-RAN), has been widely accepted to be a feasible solution for heterogeneous networks~\cite{CRAN}. In Cloud-RAN, all RAN functionalities are achieved in the centralized digital baseband units (BBUs) pool in vast warehouses of machines, which are connected to remote radio heads (RRHs) via fiber. It may take advantages of technologies such as cloud computing, advanced remote radio head (RRH) techniques and Soft-Defined Network (SDN) approaches~\cite{SDN1}. Motivated by the essence of Cloud-RAN, we propose a multi-tier cloud network architecture for HetVNET, where various resources in clouds can be jointly exploited to provide the satisfied service quality to vehicular users. Building upon Cloud-RAN, the separation between the data plane and control plane by SDN is conveniently to be implemented on the open platform~\cite{MobileSDN}. The new network architecture can improve the network efficiency while reducing the costs of management and maintenance. \par

Therefore, the \emph{\textbf{S}oft-d\textbf{E}fined hete\textbf{R}ogeneous \textbf{V}eh\textbf{IC}ular n\textbf{E}twork} (SERVICE) is expected to be a promising way for the future vehicular communication. However, there are many challenges related to the implementation of SERVICE. Although existing literatures are available for mobile SDN~\cite{OpenNB1}~\cite{OpenNB2}, to the best of our knowledge, there have been few studies reported to address the communications challenges from vehicular communication point of view. The scope of this article is thus to present the preliminary study on the cloud-based heterogeneous vehicular network. \par

The remainder of this paper is organized as follows. In Section \ref{sec_framework}, we present the SERVICE framework including a brief introduction on HetVNET and the multi-tier cloud-based architecture. Then, Section \ref{sec_sdn} presents the system design with the SDN principles in SERVICE. In Section \ref{sec_chanllenges}, we discuss networking related challenges and solutions for SERVICE. Finally, conclusions are drawn in Section \ref{Conclusions}.\par

\section{SERVICE Framework}
\label{sec_framework}
\subsection{HetVNET}

\begin{figure}
\centering
\includegraphics[width=4in, angle=270]{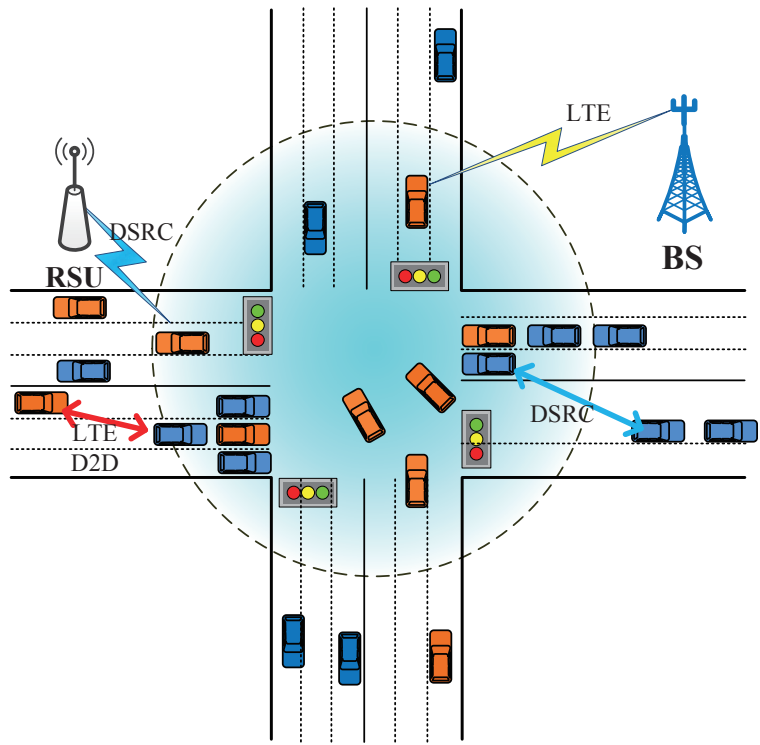}
\caption{Illustration of heterogeneous vehicular networks.}
\label{fig_HetVNET}
\end{figure}

As shown in Fig.\ref{fig_HetVNET}, different wireless access schemes, such as LTE and DSRC, co-exist together in HetVNET. Various candidate schemes can be used to support V2I and V2V communications depending on their specific characteristics. V2I communication enables the connection with the infrastructure, which is located at the roadside. LTE network has been widely deployed recently. Therefore, it is one of the most promising candidate techniques to support V2I communication. Another solution is by utilizing DSRC systems, which is designed to provide the robust, low-latency, and high throughput services for ITS. V2V communication establishes a direct connection between vehicles, minimizing traffic accident and improving traffic efficiency. Specifically, accidents caused by slow vehicles or non-sight vehicles may be avoided by exchanging their velocity, acceleration, and vehicle status to neighboring vehicles through timely V2V communication. DSRC has been proven to be very effective in supporting both safety or non-safety services in V2V communications. Meanwhile, as another candidate supporting V2V in HetVNETs, device-to-device (D2D) communication can take advantage of the physical proximity of communicating devices in LTE systems mainly for the non-safety services. Thus, to meet different QoS requirements of desired services, it is challenging to choose the suitable radio access method for each vehicular user while efficiently exploiting all the radio resources of HetVNET. \par

\subsection{Cloud-RAN Architecture}
An architecture of Cloud-RAN with the flexible soft-defined network technology is adopted for implementing the new network infrastructures for HetVNET, which conveniently integrates various wireless access schemes. It physically consists of the Remote Radio Head (RRH) and the baseband processing cloud. By using RRH, the digital or optical signals are converted to analog and amplified before being transmitted over the air, and vice versa. All the functionalities of RAN can be implemented in the baseband processing cloud. Both of them can be separately deployed and connected by the optical fiber cable. This architecture provides the feasibility to apply the open platform and real-time virtualization technology in cloud computing to achieve dynamic shared resource allocation and well support the SERVICE. Also, the SERVICE offers the flexibility for users to customize their computing environments to enjoy the desired experiences of services, i.e., ``everything as a service" or ``X as a service (XaaS)". An increasing number of services can be delivered over the Internet rather than provided locally or on-site.\par

The SERVICE is highly heterogeneous, either as a communication network or as a cloud system. From the view of communication, it consists of several kinds of base stations (BSs) with the specific wireless access techniques, e.g., either LTE or DSRC systems. Moreover, their coverage and service abilities are quite different, depending on the type of BS. For example, the macro cell BS can support tens of users in the area of several square kilometers. However, the underlaying small cells, which include femtocells, picocells, and DSRC cells, typically have much smaller coverage than macro cells. Then, the number of users supported in a small cell usually ranges no more than ten at most. On the other hand, the computing infrastructures are also heterogeneous with various processing and storage abilities. In addition, the diversity of user equipments, e.g., smart phone, pad, and smart car, enhances the network heterogeneity. \par

The architecture with multi-tier cloud is applied in SERVICE. Besides the remote cloud, the cloudlet of SERVICE is integrated with the local RAN, which provides a ``close-to-the-user" proxy system functionality that ensures timely data delivery. Physical proximity of the cloudlet is essential for the low end-to-end response time of applications, e.g., few milliseconds for safety related services. A cloudlet infrastructure consists of a cluster of communication resources and computation devices such as vehicles and servers, which are well connected and available for use by nearby users. From the perspective of a vehicle user, there are three tiers of cloud resources that it can exploit to enjoy the services, i.e.,\par

\subsubsection{Micro Cloud}
Due to rapid development of mobile computing and communication technologies, it has become feasible to use mobile nodes, e.g., smart cars, as entities in cloud services. Today many smart cars are likely to be equipped with several devices including an on-board high-performance computer with large storage, a radio transceiver, and various sensing devices, which can ensure the driving safety and enjoy the on-road multimedia services. Therefore, it is natural to regard such vehicles as ``computer-on-wheels", which consists of the bottom tier in SERVICE, called Micro Cloud (MC). The controller in the MC coordinates computing, sensing, communication and storage resources, only providing services to authorized users. Different from conventional cloud with dedicated hardware, it utilizes the available resources built in vehicles.\par

\subsubsection{Local Cloud}
The service area (SA) is defined and used as a basic geographical unit in SERVICE. The range of SAs may coincide with the coverage of macro cells, e.g., only a single or a cluster of macro cells. Each SA has its own Local Cloud (LC) with the management entity, which can jointly control not only the local communication infrastructures but also computing infrastructures. For simplicity, the resources of LC can be assumed to be deployed at the same location inside the SA, e.g., macro cell BS sites. When a vehicle user enters the coverage area of an SA, it may be connected to the nearby cloud and enjoy the services. For the sake of radio resource management, it is suggested to make use of the resources in the LC to implement the operations of the physical and media access control (MAC) layer rather than in the remote cloud. \par

Cloud-based mobile services are interactive in nature. Slow interactive responses causes a decrease in productivity and an increase in user frustration. Since the LC can be accessed directly through the wireless connection, the latency can be decreased while avoiding the wired transmission in the core networks. Receiving services from the server in the LC can always provide a better user experience. Especially in HetVNET, safety-related applications have strict latency requirements so that the corresponding service entities are allocated in the LC to ensure the instantaneity. Some kinds of cloud services may involve resources in more than one layer. In such case, the LC can act as a cache or a service proxy. Therefore, the LC plays a very important role in the multi-layer cloud architecture since it can well associate the resources in different layers.\par

\subsubsection{Remote Cloud}
For a vehicle user located in a certain SA, the remote resource pools are called the Remote Cloud (RC), which is like mobile cloud computing system. To access the RC, not only the wireless links but also the wired links are needed, which may slow the interactions between users and servers. Offloading services to RC involves the coordination between SAs, which may create a flurry of ``east-west" data traffic in wired domain. Additional signaling overheads are also unavoidable. However, the RC has powerful processing and storage capacities. When the serving ability of the LC cannot meet the QoS requirements needed by vehicles in a given SA, offloading the service request to the RC or neighboring LC becomes an good alternative.\par

\section{SDN Design in SERVICE}
\label{sec_sdn}

One of main features of SDN is the separation of control plane and data plane, and centralization of control functions. By programmable SDN controller, the network operators can easily configure new network devices and quickly deploy new applications~\cite{SDN_wp}. It is convenient to implement the soft-defined HetVNET on Cloud-RAN because of its natural character of centralized control mechanism. \par

\subsection{SDN Logical Architecture}
\label{subsec_sdn_arch}
As a soft-defined network, SERVICE can be logically divided into three layers, i.e., network infrastructure layer, control layer and application layer.  Among them, the control layer is the most important because its functions determine the behavior and performance of the network. The details of each layer are described as follows, i.e.,\par

\subsubsection{Network Infrastructure Layer}
\label{subsub_nilayer}

Network infrastructure layer is located at the bottom of SERVICE architecture, which is corresponding to the data plane of the standard SDN architecture. Actually, there are mainly several kinds of resources in this layer, i.e., the communication, computing and storage resources. The communication resources mainly consist of BBUs in clouds, RRHs and backhaul links. Through RRHs, the wireless signals can be sent or received between the BS and vehicles. The type of BSs can be flexibly adjusted according to different area features, business requirements, traffic loads, and so on. The three-tier cloud in SERVICE can provide a plenty of computing and storage resources. In order to connect all the resources, the high data rate backhaul links are needed which can be configured by the control layer. \par

\subsubsection{Control Layer}
\label{subsub_clayer}

\paragraph{Hierarchical control layer}
\label{para_hie_con}

The control layer is located at the middle of the network architecture, which acts as a service proxy translating the users' requirements and exhibits control behaviors on the virtualized resources  as well as providing relevant information to the applications. All the control functions are concentrated in the controller in traditional SDN architecture, which is suitable for centralizing management. However, due to the high mobility of vehicles, the SERVICE topology is dynamic so that the underlying network topology is instability. The amount of information exchanged between different layers and functions becomes huge so that the real-time communication performance is hard to be guaranteed. Therefore, the centralized controller maybe becomes inefficient in such cases. Moreover, when the network scale is enlarged, the complexity of functions in a single controller is increased too rapidly to be realized. Then, a more effective way is desired to make QoS guaranteed in SERVICE, which can deal with the quick response on requests from fast-moving vehicles. Thus, we propose a hierarchical control layer in SERVICE, which consists of the Primary Controller (PCon) and the Secondary Controller (SCon).\par

\begin{itemize}
\item{\emph{Primary Controller:}}
\label{subpara_prim}
Primary controller (PCon) lies on the top layer of the SERVICE control layer. PCons are responsible for controlling the global SERVICE network. Generally, PCons include wide area or non-real-time control functions, such as inter-SA handoff, cloud resource allocation, and so on. Global network information is assembled in PCons, such as control layer topology, SAs states and resources states. With these knowledge, PCons may make the possible optimal decisions. Furthermore, PCons have interface with the network infrastructures directly as well as to SCons. \par

\item{\emph{Secondary Controller:}}
\label{subpara_secn}
SCons are located logically below PCon, which play an important role as a regional control entity. The effective areas of SCons are based on the SAs, i.e., each SCon controls one SA. One of main responsibilities of SCons is to assure the QoS requirements of low-latency safety applications, e.g., V2V communication control. Also, SCons control the intra-SA resources through managing a virtual resource pool. \par
\end{itemize}

\paragraph{Controller interface}
\label{para_ci}

The principle of SERVICE is to separate the different layers so that there is no need for applications to know anything about underlying devices, and vice versa. All the communications between bottom and top layers are achieved via different controller interface in the middle layer. There exist three kinds of interfaces in SERVICE controller, i.e.,\par

\begin{itemize}
\item{\emph{North-Bound Interface:}}
\label{subpara_nbi}
SERVICE controller North-Bound Interface (NBI) is divided into two types, i.e., North-Bound Interface-Application (NBI) and North-Bound Interface-Primary Controller (NBI-P). NBI is a set of standard and open APIs, providing communications between controllers and applications. With NBI, applications can exchange data with controllers. Controllers provide an abstract network infrastructures for applications via NBI, thus SERVICE administrators can design applications without any information of underlying infrastructures. NBI-P is needed for SCons to communicate with PCons in the Hierarchical control layer of SERVICE. All the network information is sent to PCons from SCons via NBI-P.\par

\item{\emph{South-Bound Interface:}}
\label{subpara_sbi}
There are two types South-Bound Interface, i.e., South-Bound Interface-Infrastructures (SBI-I) and South-Bound Interface-Secondary controllers (SBI-S). Due to the concentration characteristic of control functions in SERVICE, controllers need to communicate with the underlying infrastructures by sending control commands via SBI-I. In fact, all the information exchanged between controllers and network infrastructures go through SBI-I, including various network states that are measured by underlying devices. SBI-S only exists in PCons, which is helpful for the communications from PCons to SCons. With SBI-S, SCons can act as a bridge connecting PCons with infrastructures. \par

\item{\emph{West-East-Bound Interface:}}
\label{subpara_webi}
As SERVICE network scale may become extremely large both geographically and logically, only a single controller cannot well handle the huge control load, especially for PCon. To avoid the high risk of overloading, SERVICE controllers are designed to be extensible. Each controller has not only West-Bound Interface (WBI) but also East-Bound Interface (EBI). These interfaces make controllers connected horizontally. With them, PCon can help to each other to balance the control loads in the entire network, which can avoid the overload problem. Moreover, SCons may form the local group and deal with the control tasks together under the coordination of PCon. In addition, it becomes much easier to deploy new controllers in SERVICE if necessary. \par

\end{itemize}

\paragraph{Classification of Control Layer Functions}
\label{para_class}

Control layer exerts its influence by different control functionalities. Generally, all control functionalities in SERVICE can be divided according to their usages, i.e., \par
\begin{itemize}
\item {\emph{Communication Control:}}
The communication resources are managed by the communication control functions. Main responsibilities of these functions include the coordination and control of physical and MAC operations for the heterogenous network infrastructures and devices of vehicle users. For example, in case of deciding how to choose the suitable way of V2V communication,  the decision application require the aids of control functions such as \emph{Network Discovery Control}, \emph{Topology Detection Control}, and so on. \par
\item {\emph{Computing Control:}}
There is another kind of control functions mainly for cloud computing operations, such as \emph{Resource Sensing Control}, \emph{Resource Virtualization Control}, \emph{Offloading Control}, and so on.\par
\item {\emph{Storage Control:}}
To minimize the SERVICE cost as well as meet the requirements of the applications, the storage control functions can be classified into two types. One of them is to guarantee the demands of the applications, e.g., \emph{Storage Capacity Control}. This type of functions configure the storage infrastructure capabilities based on the application requirement. Another is to save the system cost. For example, the function of \emph{Data Migration Control} can move the data that not requested frequently to the low cost storage devices~\cite{storage}.\par

\end{itemize}

\subsubsection{Application Layer}
The application layer is the top layer of SERVICE, by which operators can configure and control SERVICE via designing different applications, e.g.,\par
\begin{itemize}
  \item{\emph{Access Management:}}
  When the Access Management application runs, the SCons monitor the network load and radio link status. Once the network load in a given vBS exceeds a certain level, SCons control the new vehicle access requests to another vBS, not only achieving the traffic load balance as well as meeting the QoS requirements of existing vehicle users.\par
  \item{\emph{Dynamic Resource Allocation:}}
   In SERVICE, each vehicle is treated as a small resource unit. SCons collect all the information of different kinds of resources into its virtual resource pool.  All resources in the total network can be optimized by dynamic resource allocation application. Once there comes new service request, this application may find a way to meet it according to current network states. After interaction with applications through NBI, controllers are responsible to allocate available resources to users in guidance of applications.\par

\end{itemize}

\subsection{Concept of Virtual Base Station (vBS)}
Based on the above architecture, soft-defined virtual base stations (vBSs) can be implemented to offer the qualified services to vehicle users. Basically, one vBS consists of virtualized resources mainly in the LC and the corresponding connected RRHs. The type communication of a given vBS is determined and configured by PCon. For example, a macro cell type of vBS, i.e., MvBS, is used in order to meet the coverage requirement within a single SA. On the other hand, the services with either high data rate or short latency can be provided by small cell type of vBS, i.e., SvBS. Since the implementation complexity and communication ability are quite different between MvBS and SvBS, more resources are usually allocated to MvBS than to SvBS in SERVICE. \par

Corresponding to the layered-architecture, there are three layers in each vBS of SERVICE. The physical infrastructure in the data plane is abstracted and handled by the functions in the control plane. For example, the computation resources in the data plane can be utilized with the unit of virtual machine (VM). Generally the vBSs have all necessary functionality to deal with the allocated physical resources and realize the data/control plane towards vehicle users.  A variety of applications can be deployed in the application layer to ensure the vBSs work properly and possibly be optimized. In the high-density deployment area, an interference coordination application is needed in the neighboring vBSs for avoiding the inter-cell interference. \par

\subsection{Benefit of SDN in SERVICE}

Since the network infrastructure and radio resources in SERVICE can be easily shared among different network operators, the capital expenses (CAPEX) and operated expenses (OPEX) of deploying new ITS services can be decreased significantly. Meanwhile, different services can be well co-existed in the heterogenous network. Also, the logically centralized control on the network infrastructure and resources gives the opportunity to maximize the system utility. Moreover, regarding to the strict QoS requirement services, e.g., safety-related service, the controller can timely allocate dedicated resources to guarantee the short latency.\par

\section{Challenges and Solutions in SERVICE}
\label{sec_chanllenges}

The use of SDN in HetVNET raises a number of new challenges in the network design. In this section, we first present an overview of the important control functions in SERVICE, i.e., \emph{Multi-domain Resource Virtualization}, which is the essential for network control and management.  Then, several key applications are presented from the view of resource optimization.  After being abstracted, resources can be efficiently scheduled either in homogenous network or in the heterogeneous network by the application of \emph{Cooperative Wireless Scheduling}. Moreover, by applying \emph{Joint Utilization} application, the multi-domain resources can be efficiently exploited together with the trade from one to another. Moreover, to reduce the handover outage in the ultra-dense deployment, the \emph{Mobility Management} application is implemented at the MvBSs for the purpose of seamless connectivity.
\par

\subsection{Multi-domain Resource Virtualization}
\label{sec_virtual}
Since the resources in SERVICE exist in different domains including communication, computation and storage domain, it is necessary to virtualize all of them uniformly in order to manage them flexibly and efficiently. The control layer in SERVICE can shelter from the differences of the multi-domain resource (MDR) in the infrastructure layer by virtualization and forming different types of vBSs by scheduling the resources as required by the network. Then, several vBSs in one or multiple SAs can be organized to be the virtual radio access network (vRAN) with necessary functions, which is in accordance with the service requirement and the load capacity that vBSs can provide. With the use of the access control application, the vehicle users can connect to one of vRANs and make use of the MDR.\par

\begin{figure}
\centering
\includegraphics[width=4.5in, angle=270]{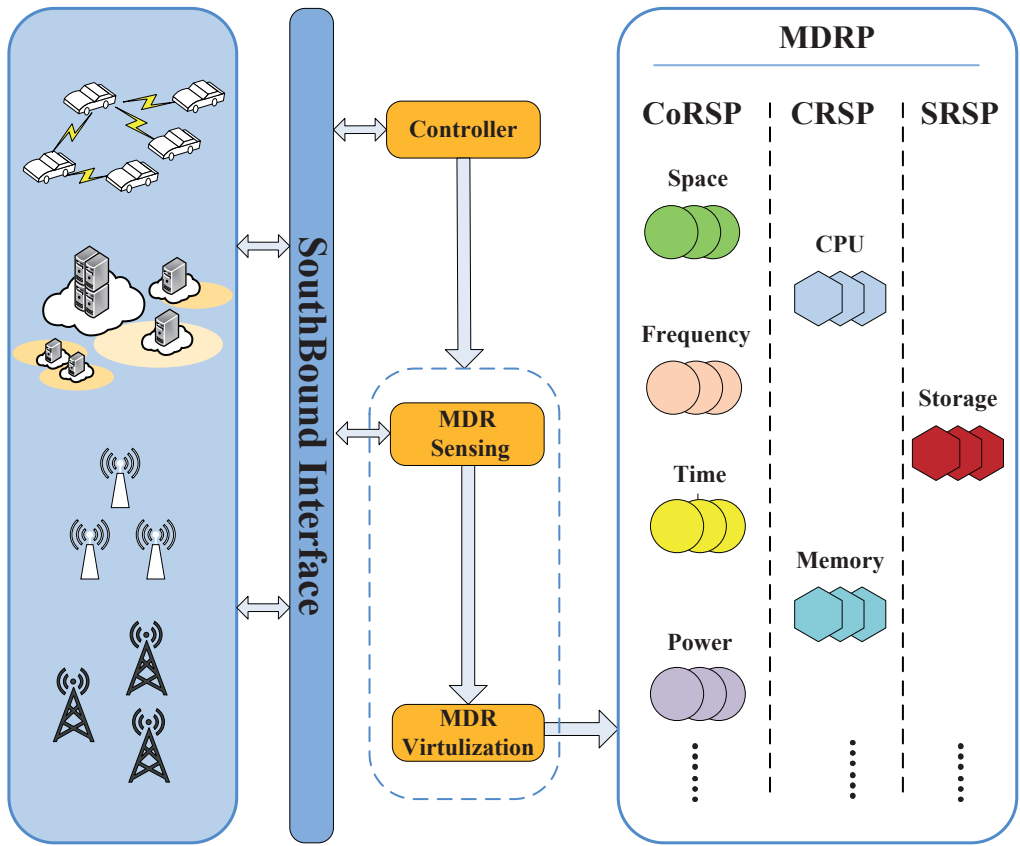}
\caption{MDR sensing and virtualization in SERVICE.}
\label{fig_emp1}
\end{figure}

One of the main challenges to exploit the MDR is how to abstract and manage communication, computation and storage resources. As shown in Fig.\ref{fig_emp1}, all the MDRs are processed and mapped to the MDR pool (MDRP) by MDR virtualization entity after being fully sensed through MDR sensing (MDRS) entity. All of these processes are coherently controlled in order to rapidly reuse the MDR in SERVICE.\par

\subsubsection{Multi-Domain Resource Sensing (MDRS) Entity}
\label{subsub_mdrs}
Due to the dynamic characters of resources, it is essential to sense the MDR timely and accurately. For example, the capacity provided by the communication resources depends on the time-varying wireless channels. Moreover, the spectrum usage in different areas are not instant, which means the unused spectrum resources in a given place may be exploited by another place if being found in time. Meanwhile, the computation and storage resources in the tiers of cloud are also not invariant due to instability of vehicles in MC or sudden failure of servers in cloud. Therefore, the MDRS entity in the control layer plays a critical role in SERVICE to quickly find and utilize the multi-domain resources. There are mainly two kinds of sensing methods, which are categorized by how much and how often resource information reported to the controller, i.e.,\par

\begin{itemize}
    \item{\emph{Regular sensing:}}
    \label{item_reg}
    The information of available MDR is sensed and reported periodically to the Multi-Domain Resource Virtualization (MDRV) Entity. Correspondingly, the MDRV records and updates the information of MDR in MDRP. However, the sensing frequency has to be determined by balancing between the system overhead and performance. The Controller can adjust the sensing period of the MDRS entity based on the network environment to reduce the sensing cost, e.g., shortening the sensing period when the network environment changes slowly. Moreover, the controller can configure the different sensing periods for different areas in the SA so as to make it more efficiently.\par

    \item{\emph{Event-evoked sensing:}}
    \label{item_eve}
    If an accident or emergency happens in the network such as the interrupt of a certain link or traffic overload in the given SA, the warning message needs to be reported to the controller immediately by event-evoked sensing.\par
\end{itemize}
Through these sensing methods, the information of MDRP can be maintained accurately, which is of paramount importance for cooperative allocation of MDR.\par

\subsubsection{Multi-Domain Resource Virtualization (MDRV) Entity}
\label{subsub_mdrv}

Through MDR virtualization, all the MDR in SERVICE are categorized and virtualized to generate the MDRP. The pool is not physically existed but only contains the logical represented information mapped from the physical resources. Different from exclusively possessing some kinds of given resources in the traditional networks, all the vBSs in SERVICE can share all the MDR in MDRP under the management of the controller. Corresponding to the existed resources in the SERVICE infrastructure layer, there are three sub-pools in MDRP, i.e.,\par

\begin{itemize}
    \item{\emph{Communication resource sub-pool (CoRSP):}}
    \label{item_corsp}
    The communication resources in CRSP are abstracted and represented by a multiple dimensional grid of space, time, frequency and others, e.g., ``vBS index, space, time, frequency" for the resources in local and RC and ``vehicle index, space, time, frequency" for the resources in the MC. All the grid information is saved in the database and updated as needed.\par

    \item{\emph{Computation resource sub-pool (CRSP):}}
    \label{item_crsp}
    CRSP has the virtualized information of all the computation resources in the three ties of cloud in SERVICE. Due to the different investment and deployment in sites, the cloud computation ability of different places cannot be identical, which means LC and RC are heterogeneous. Furthermore, vehicles produced by different manufactures, naturally have the different computation capabilities, which is the base unit of MC. Therefore, to be jointly exploited, these heterogeneous resources are needed to be virtualized into a uniform form, i.e., virtual machine, without concerning on hardware details. Then, each computation resource may be labeled as the grid like ``VM index, CPU, memory".\par

    \item{\emph{Storage resource sub-pool (SRSP):}}
    \label{item_srsp}
    Similarly, abstraction of heterogeneous resources is also of vital importance for future unified management. The storage resources distributed in the tiers of cloud can be integrated to develop the virtual storage pool so as to make them transparent to the applications in the application layer. Moreover, the SRSP can further be decomposed into multiple capacity pools, which are organized by geography or service class~\cite{storage}. Furthermore, in order to allocate the resources conveniently, the storage resources can be abstracted into consumable units of a block, file, or object type. Through the virtualization, the storage resource utilization and the operation efficiency can both be improved.\par

\end{itemize}

It is essential to divide the MDR into the unified resource units (RUs) in order to let all the resources in different clouds be used with the comparable metric. If the size of one RU is too small, it is hard to find the feasible way to measure the same kind of resources in different systems, especially for the communication resource in heterogeneous networks. Meanwhile, the smaller RU granularity means the more flexible for performance optimization but the higher control cost, vice versa. Furthermore, since not only the resource usage but also traffic load vary rapidly, the fixed size of RU is not the optimal choice for the sake of the system performance and may be adjusted adaptively. Therefore, to select the proper RU granularity in CoRSP, CRSP and SRSP is the crucial step to guarantee the performance of SERVICE.\par

\subsection{Cooperative Management on Wireless Resource}
\label{sub_coop}

Due to the high speed of moving vehicles, vehicles frequently move into the non-coverage areas of vBSs, such as in the tunnel. Also, vehicle user experiences low data rate when traveling at the edge of vBS. Then, the cooperative communication is necessary to improve the traveler's experience during the movement. Based on various access technologies integrated in SERVICE, ``Cooperative Management" applications can be roughly classified into heterogeneous and homogeneous cooperation. \par

\subsubsection{Homogeneous Cooperation}
When vehicles are mainly served by a single wireless access system in a specific area, only the homogeneous cooperation is needed.  One of the important cooperative objectives in LTE systems is how to eliminate the interference from neighboring vBSs. The inter-cell interference coordination (ICIC) can be carried out in either of the space, frequency and time resource dimension. The space dimensional ICIC is to concentrate the transmitted signal to the desired user while reducing the interference to other users via beamforming technique. By configuring different priorities on the same frequency resources of the neighboring vBSs, the inter-cell interference can be efficiently mitigated. On the other hand, in the Almost Blank Subframe (ABS) scheme, different frames are allocated to different users to avoid harmful the interference in the time dimension. Among all of the above ICIC schemes, it may be beneficial to have the inter-cell information as much as possible.  The centralized architecture of SERVICE provides the feasibility of exchanging such information, e.g., more than one vBS are located in a same LC and controlled by a single SCon. The overhead and latency for messages between vBSs may not be a serious concern in such a network. \par

\subsubsection{Heterogeneous Cooperation}
\begin{figure}
\centering
\includegraphics[width=7in]{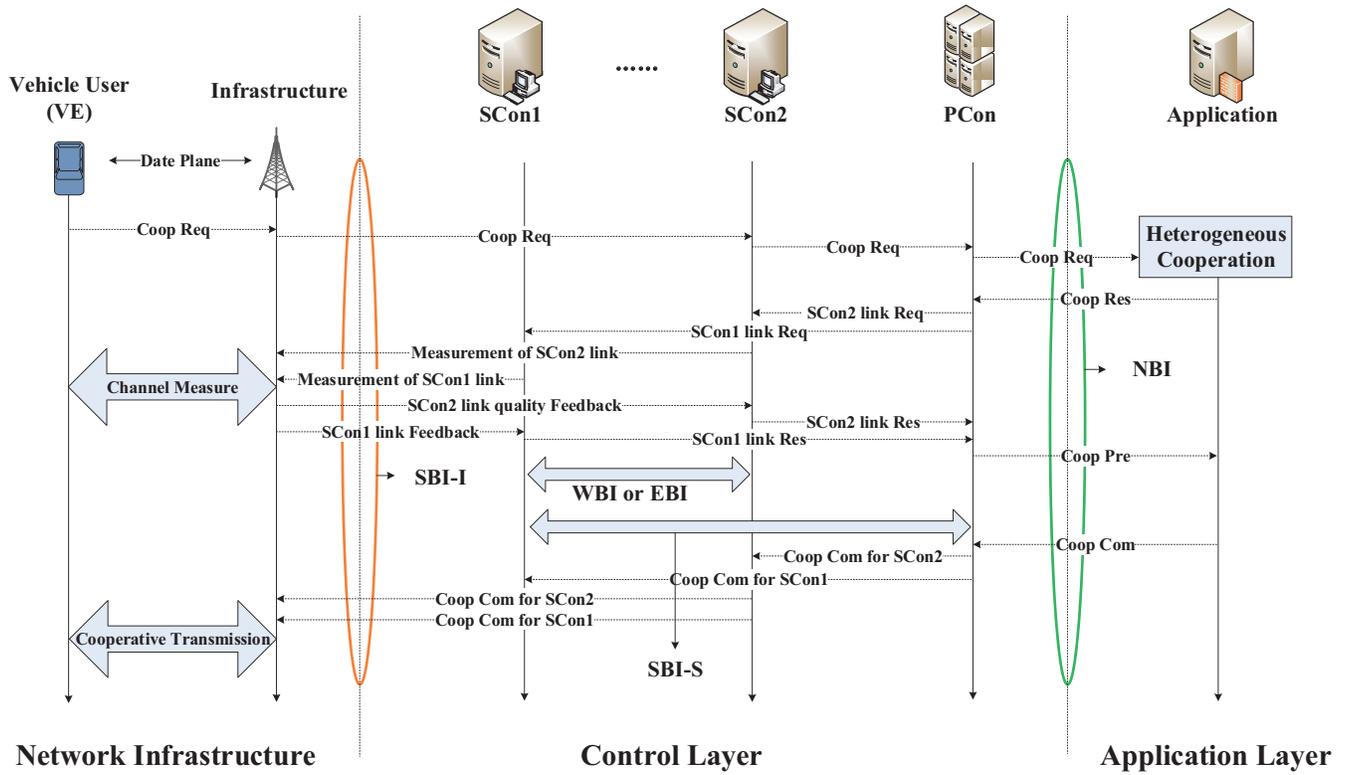}
\caption{An example of heterogeneous cooperative transmission in SERVICE.}
\label{fig_het_coop}
\end{figure}

Heterogeneous cooperation usually involves at least two different wireless access system, e.g., LTE and DSRC. When a vehicle detects that the data rate of LTE link is below certain threshold or the vBS detects the failure of communication, the heterogeneous cooperation may be triggered. An example of using DSRC to improve the LTE communication quality is shown in Fig.\ref{fig_het_coop}. Firstly, the vBS that the vehicle is currently associated sends the \emph{Coop Req} message to the SCon via SBI-I. Then, the SCon analyzes the request and forwards it to the PCon, which communicates with the application of ``Heterogeneous Cooperation" via NBI. In this way, the application is executed and sends the feedback of \emph{Coop Res} message to PCon. After parsing the message successfully, the PCon requests the necessary information, e.g., the Channel State Information (CSI) of the surrounding LTE links and V2V links. After received all necessary information from controllers, a corresponding heterogeneous cooperative scheme between LTE and DRSC systems is carried out with the control messages issued by controllers. There may exist different schemes with different optimization objective, e.g., either minimizing the delay/outage probability, or maximizing the throughput. \par

\subsection{Joint Utilization on Multi-Domain Resource}
\label{subsub_coop}

Thanks to the high-performance computation and massive storage techniques, the SERVICE ability is not only dependent on the communication infrastructures but also the computation and storage devices in the network. In other words, the good service quality to a vehicle user can be achieved by trading one kind of resource with another. Therefore, one kind of key applications in SERVICE is how to allocate multi-domain resources in the MDRP in order to provide the satisfied QoS for vehicle users with the low network cost.\par

There are typically two ways to get the equivalent communication performance gain at the cost of either storage or computation resources, i.e.,\par

\begin{itemize}
    \item{\emph{Storage cost:}}
    \label{item_sto_cost}
    Mass storage can be deployed in LC, which is very close to vehicle users. The most frequent used contents such as local traffic and road conditions, and popular news maybe stored in LC and broadcast to the nearby vehicle users. Then, the end-to-end latency between vBS and vehicle users can be reduced. However, more storage capacity is needed due to the increase of common-interested contents with the increase number of vehicle users.\par

    \item{\emph{Computation cost:}}
    \label{item_comp_cost}
    Before being delivered to the storage and transmission, the service data can be first analyzed with the cost of computation resources. According to the analysis on the dimensions such as location, activity and interest, vehicle users can be classified into tribes with the specific characters. Based on the tribe features, the controller allocates communication resources with the proper transmission to vehicle users. For example, the information to users in the same tribes preferred to be transmitted by broadcasting rather than unicasting. Then, communication resources can be saved for other users. If possible, the broadcast range may cover more than one SA in order to further reduce the requirements of communication resources in the given area.\par

\end{itemize}

On the other hand, the storage and computation ability of vehicle users can also be improved somehow at the cost of communication resources. The storage volume can be enlarged by transmitting the data of vehicle users to the tiers of cloud. The computation power can also be enhanced by offloading the computing task to the cloud whose processing ability is much more powerful. However, whether the performance of vehicle users is improved or not depends on the communication capacity, which is instable and varies frequently. If the vehicle user utilizes the cloud at inappropriate time, it may lead to too much energy consuming and too large latency. Offloading is a typical application example, which decides whether the computing task at vehicle user is offloaded to the cloud based on the communication and computation resources allocated.\par

As the improvement on batteries is not match with the rapid growth in wireless traffic, offloading is regarded as a possible way to overcome the obstacle. Whether the energy can be saved or not by offloading depends on the wireless bandwidth and the processing power of the cloud server, i.e., the computation resources allocated to the vehicle user, given the amount of computation to be performed and the data to be transmitted~\cite{offloading}. Therefore, it is necessary to jointly allocate the communication and computation resources in the cloud. The decision procedures in SERVICE can be shortly described as follows, i.e.,\par

\begin{itemize}
    \item{\emph{Step 1:}}
    \label{item_step_1}
    The vehicle user triggers an offloading request and the Controller sends the request to the Offloading Application in the application layer.\par

    \item{\emph{Step 2:}}
    \label{item_step_2}
    The \emph{Offloading Application} asks for not only the relevant information of the MDR in MDRP but also the related information of the request.\par

    \item{\emph{Step 3:}}
    \label{item_step_3}
    Based on the information received, the \emph{Offloading Application} makes the decision of whether accepting the offloading request or not. If the request is accepted, the decision of accessing to which vBS and how to allocate the multi-domain resources to the request is made to maximize the revenue of SERVICE.\par

\end{itemize}

Since the dynamic arrive character of the offloading requests and the instability of the resources in MDRP, the steady optimal algorithm, which makes decision just based on the present global state, may no longer meets the requirements to maximize the system revenue. So, the dynamic programming algorithm is necessary to solve the offloading problem. Possibly, this optimal problem can be formulated as a Markov decision progress (MDP), i.e.,\par

\begin{itemize}
    \item{\emph{State:}}
    \label{item_state}
    the available communication resources of each vBS, the channel state between vehicle user and vBSs, the channel state between vehicle users, the available computation resources of the three tiers of cloud.\par

    \item{\emph{Action:}}
    \label{item_action}
    Whether accept the offloading request; if accept, make a decision of accessing to which vBS and allocate corresponding communication and computation resources.\par

    \item{\emph{Reward:}}
    \label{item_reward}
    The revenue obtained by the SERVICE. The revenue may be different based on the special objectives, e.g., minimum transmit energy under computational constraint, minimum transmit power under delay constraint~\cite{cc}, minimum both transmit energy consumption and delay~\cite{leil} and so on. \par

    \item{\emph{Transition Probability:}}
    \label{item_prob}
    The transfer of each channel state, the available communication resources in each vBS and available computation resources in each tier.\par

\end{itemize}

Since resources in different domains can be traded by each other, we can allocate the multi-domain resources cooperatively so as to optimize the performance of SERVICE. It is fortunate that the centralized control architecture in SERVICE facilitates the implementation of such allocation.\par

\subsection{Mobility Management in SERVICE}
\label{sub_mob}

Due to the limited coverage of a single base station, the handover may be happened frequently when vehicle users travel with the high mobility, either in LTE or in DSRC. On the other hand, since the radio resources can be geographical reused in different SAs, their virtualization and management are usually responsible by a SCon in the local SA. So, there exists four kinds of handover in HetVNET, i.e.,\par
\begin{itemize}
    \item{}
    \label{item_iapias}
    Intra-SA-Intra-System Handover: The vehicle user changes connection between two vBSs, which works in the same wireless access technique in the same SA, e.g., the handover between MvBS and SvBS in LTE.\par
    \item{}
    \label{item_irpias}
    Inter-SA-Intra-System Handover: This situation occurs when the vehicle user moves from one SA to another but are served by the same wireless access system.\par
    \item{}
    \label{item_iapirs}
    Intra-SA-Inter-System Handover: The vehicle user changes its serving node from one wireless access system to another; both of them operate in the same SA.\par
    \item{}
    \label{item_irpirs}
    Inter-SA-Inter-System Handover: The vehicle user has to associate to another wireless access system when it moves from one SA to another. \par

\end{itemize}

By applying the separation of data and control channel into SERVICE,  we can reduce the possibility of the handover happened. The main ideology is that only the MvBS provides mobility-related control channel for its reliable and continuous coverage while the SvBS is responsible for high data rate transmission to achieve the ``\textbf{Wi}de Co\textbf{n}trol and \textbf{Lo}cal \textbf{T}ransmission\textbf{s}",  termed as ``\textbf{Win-Lots}". In other words, there is no need to handover if a user moves within the coverage of one SA. It can break the bottleneck between coverage and transmission capacity.  The radio resources on the low frequency band are preferred to be used by the MvBSs due to their low propagation path loss. On the other hand, the high frequency bands with a plenty of radio resources are more likely to be allocated for the SvBS. Based on this principle,  ``Mobility Management" applications can be carefully designed to minimize the outage probability during handover.\par

By using storage resources in cloud, the Vehicle Mobility Monitoring (VMM) function in the PCon records and updates the trajectory of vehicles. Meanwhile, vehicle user can report the location information to VMM via MvBS control channel.  ``Mobility Management" applications may be activated by the controller in certain scenarios, e.g., when a vehicle user is crossing the edges of the neighboring SAs. Then, the command messages are sent to the control functions. \par

There are still several issues that need to be concerned in \textbf{Win-Lots mode}. For example, since the control channel has been simplified in SvBS, a new control channel format has to be designed without the cell-specific signaling. \par

\section{Conclusion}
\label{Conclusions}
The soft-defined HetVNET is very promising to meet the requirements of future vehicular communications. A plenty of multi-domain resources can be virtualized and pooled in clouds, which can be exploited by vehicle users to enjoy either safety-related or non-safety-related services. In this paper, we have investigated the feasibility of SDN in HetVNET based on Cloud-RAN architecture, called as SERVICE. A new hierarchical control layer is also proposed to facilitate the SERVICE implementation. It can provide the flexibility, centralized control, and open interfaces between functions and layers, enabling a unified and flexible network. However, lots of challenges have to be tackled before it comes true. Although we have presented the preliminary study on the key control functions and applications, much more investigation is needed in order to make them into practice. Nevertheless, it is certain that the soft-defined HetVNET may play an important role towards a ubiquitous 5G network.\par

\section*{Acknowledgment}
This work is funded in part by the National High-Tech R\&D Program (863 Program 2015AA01A705), National Science Foundation of China (No.61331009), National Key Technology R\&D Program of China (No.2015ZX03002009-004) and Fundamental Research Funds for the Central Universities (No.2014ZD03-02).

\section*{Biography}

\begin{small}

\textbf{Kan Zheng}
(SM'09) is currently a full professor in Beijing University of Posts \& Telecommunications (BUPT), China. He received the B.S., M.S. and Ph.D degree from BUPT, China, in 1996, 2000 and 2005, respectively. He has rich experiences on the research and standardization of the new emerging technologies. He is the author of more than 200 journal articles and conference papers in the field of wireless networks, M2M networks, VANET and so on. He holds editorial board positions for several journals. He has organized several special issues in famous journals including IEEE Communications Surveys \& Tutorials, IEEE Communication Magazine and IEEE System Journal.

\textbf{Lu Hou}
is now a candidate for M.S. in the Key Lab of Universal Wireless Communications, Ministry of Education, BUPT. He earned his B.Eng. degree from the School of Information and Communication Engineering, Beijing University or Posts and Telecommunications (BUPT), China, in 2014. Nowadays, he mainly focuses on Soft-define Network (SDN) and resource allocation in mobile cloud computing.

\textbf{Hanling Meng}
received her B.Eng. degree from the School of Automation, Beijing University or Posts and Telecommunications (BUPT), China, in 2013. She is currently a candidate for M.S. in the Key Lab of Universal Wireless Communications, Ministry of Education, BUPT. Her research interests include mobile cloud computing (MCC), vehicular network, software-defined network (SDN) and corresponding resource allocation.

\textbf{Qiang Zheng}
received his B.S. degree from the College of Computer Science and Technology, Shandong University of Technology (SDUT), China, in 2010. He is currently a Ph.D. candidate in the Key Lab of Universal Wireless Communications, Ministry of Education, Beijing University of Posts and Telecommunications (BUPT). His research interests include radio resource allocation, performance analysis, and optimization in heterogeneous vehicular networks.

\textbf{Ning Lu}
(S'12) received the B.Sc. (2007) and M.Sc. (2010) degrees from Tongji University, Shanghai, China, and Ph.D. degree from University of Waterloo, Waterloo, ON, Canada, all in electrical engineering. He is currently working as a postdoctoral fellow with the Department of Electrical and Computer Engineering at the University of Waterloo. His research focuses on fundamentals of wireless networking with special interest in connected vehicles. Mr. Lu served as a Technical Program Committee Member for IEEE PIMRC'12, WCSP'13, WCSP'14, and ICNC'15.

\textbf{Lei Lei}
received a B.S. degree in 2001 and a PhD degree in 2006, respectively, from Beijing University of Posts \& Telecommunications, China, both in telecommunications engineering. From July 2006 to March 2008, she was a postdoctoral fellow at Computer Science Department, Tsinghua University, Beijing, China. She worked for the Wireless Communications Department, China Mobile Research Institute from April 2008 to August 2011. She has been an Associate Professor with the State Key Laboratory of Rail Traffic Control and Safety, Beijing Jiaotong University, since Sept. 2011. Her current research interests include performance evaluation, quality-of-service and radio resource management in wireless communication networks.

\end{small}

\clearpage

\clearpage
\begin{figure}
\centering
\includegraphics[width=7in]{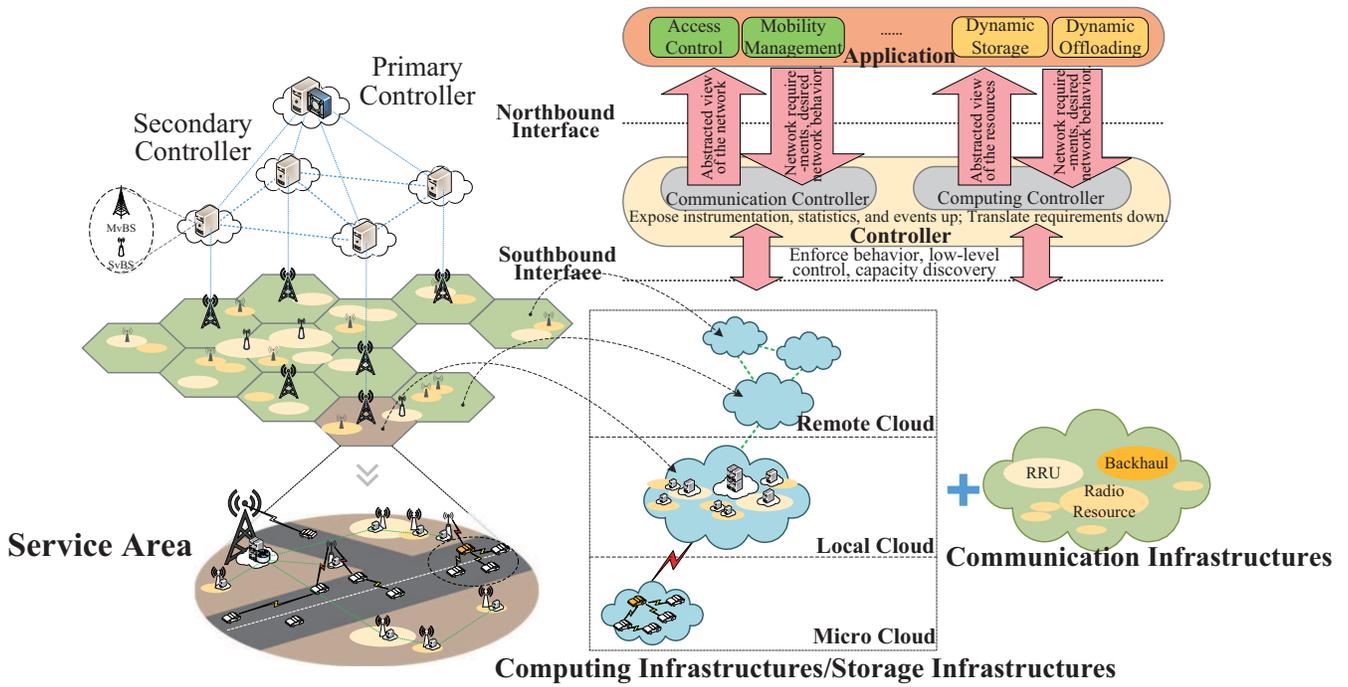}
\caption{Illustration of Cloud-based architecture in SERVICE.}
\label{fig_hie}
\end{figure}

\clearpage
\begin{figure}
\centering
\includegraphics[width=7in]{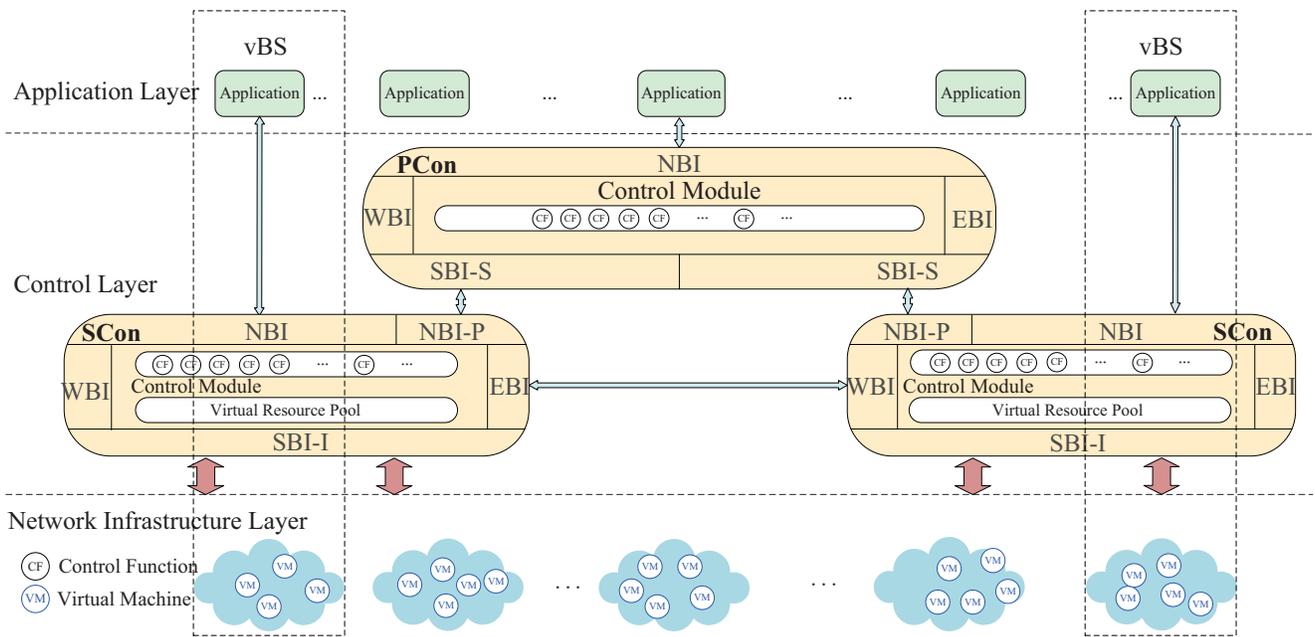}
\caption{Hierarchical Control Layer in SERVICE.}
\label{fig_hie}
\end{figure}

\clearpage

\clearpage

\end{document}